# The Human Body and Millimeter-Wave Wireless Communication Systems: Interactions and Implications

Ting Wu[1,2], Theodore S. Rappaport[1,2,3], Christopher M. Collins[1,3]
[1]NYU WIRELESS, [2]NYU Polytechnic School of Engineering, Brooklyn, NY 11201
[3]NYU Department of Radiology, 660 First Ave, New York, NY 10016
ting.wu@nyu.edu, tsr@nyu.edu, c.collins@nyumc.org

*Abstract*—With increasing interest in millimeter-wave wireless communications, investigations on interactions between the human body and millimeter-wave devices are becoming important. This paper gives examples of today's regulatory requirements, and provides an example for a 60 GHz transceiver. Also, the propagation characteristics of millimeter-waves in the presence of the human body are studied, and four models representing different body parts are considered to evaluate thermal effects of millimeter-wave radiation on the body. Simulation results show that about 34% to 42% of the incident power is reflected at the skin surface at 60 GHz. This paper shows that power density is not suitable to determine exposure compliance when millimeter wave devices are used very close to the body. A temperature-based technique for the evaluation of safety compliance is proposed in this paper.

## I. Introduction

THE millimeter-wave (mmWave) band is part of the radio frequency (RF) spectrum, comprised of frequencies between 30 GHz and 300 GHz, corresponding to a wavelength range of 10 to 1 mm. The photon energy of mmWaves ranges from 0.1 to 1.2 milli-electron volts (meV). Unlike ultraviolet, X-ray, and gamma radiation, mmWave radiation is non-ionizing, and the main safety concern is heating of the eyes and skin caused by the absorption of mmWave energy in the human body [1][2][3]. The massive amount of raw bandwidth and potential multi-Gigabit-per-second (Gbps) data rates in the mmWave band make it a promising candidate for future broadband mobile communication networks [3][4]. The increasing investigations on mmWave applications and technologies, particularly on wireless devices, have stimulated interest in understanding how propagation of mmWaves impact the human body, as well as the inquiry of potential health effects related to mmWave exposures.

MmWave devices should be evaluated to comply with government exposure guidelines before they are introduced to the consumer market. At frequencies below 6 GHz for the Federal Communications Commission (FCC) or 10 GHz for the International Commission on Non-Ionizing Radiation Protection (ICNIRP), the specific absorption rate (SAR) is used as a metric for exposure compliance determination. However, at higher frequencies, energy absorption is increasingly confined to the surface layers of the skin, and it is difficult to define a meaningful volume for SAR evaluation. Thus, power density (PD), rather than SAR, is currently preferred in determining compliance at above 6 GHz (FCC) or 10 GHz (ICNIRP) [1][2][3].

The ICNIRP specifies basic restrictions on PD to be 10 W/m$^2$ and 50 W/m$^2$ for the general public, and the occupational group, respectively, for frequencies between 10 and 300 GHz [1]. The limit values are to be averaged over any 20 cm$^2$ of exposed area and any $68/f^{1.05}$ minutes period (where $f$ is in GHz), while the spatial peak power densities averaged over 1 cm$^2$ should not exceed 20 times the given limits, which are 200 W/m$^2$ and 1000 W/m$^2$, respectively.

The FCC adopts maximum permissible exposure (MPE) in terms of PD for frequencies between 6 and 100 GHz [5]. The numerical values of the FCC PD restrictions are also 10 W/m$^2$ and 50 W/m$^2$ for the general public, and occupational group, respectively, while the exposure area to be averaged for the FCC is equivalent to the vertical cross section of the human body (projected area) at a distance no closer than 20 cm from the field source. The averaging time is 6 minutes for occupational exposures, and 30 minutes for general population exposures.

Regarding localized peak power density, FCC OET Bulletin No.65 [6] states that "although the FCC did not explicitly adopt limits for peak power density, guidance on these types of exposure can be found in Section 4.4 of the ANSI/IEEE C95.1-1992 standard." The ANSI/IEEE C95.1-1992 standard specifies relaxation of PD limits for exposure of all parts of the body except the eyes and the testes [7]. For frequencies between 3 and 15 GHz, the averaging time is 90,000/$f$ (where $f$ is in MHz), and for frequencies between 15 and 300 GHz, the appropriate averaging time is 616,000/$f^{1.2}$ minutes (where $f$ is in MHz). For occupational/controlled exposures, the peak power density should not exceed $200(f/6)^{1/4}$ W/m$^2$ at frequencies between 6 and 96 GHz (where $f$ is in GHz), and 400 W/m$^2$ at frequencies between 96 and 300 GHz. For general population/uncontrolled exposures, the peak PD should not exceed $10(f/1.5)$ W/m$^2$ for frequencies between 6 and 30 GHz ($f$ is in GHz), and 200 W/m$^2$ at frequencies between 30 and 300 GHz.

While the FCC has not updated the statements regarding limits on peak power density for localized exposure scenarios issued about 20 years ago, the ANSI/IEEE C95.1 standard has been modified with the evolution of technology. In the ANSI/IEEE C95.1-2005 standard, relaxation of the PD MPEs is allowed for localized exposures on any part of the body [2]. The PD are intended to be spatially averaged over an area of $100\lambda^2$ for frequencies below 30 GHz ($\lambda$ is in cm), and averaged over 100 cm$^2$ for frequencies above 30 GHz. The averaging time is 6 minutes for occupational/controlled exposures, and 30 minutes for general population/uncontrolled exposures. For exposures in controlled environments, the spatial peak value of the PD shall not exceed $200(f/3)^{1/5}$ W/m$^2$ at frequencies between 3 and 96 GHz ($f$ is in GHz), and 400 W/m$^2$ at frequencies from 30 GHz to 300 GHz. For exposures in uncontrolled environments, the spatial peak value of the PD shall not exceed $18.56(f)^{0.699}$ W/m$^2$ at frequencies between 3 and 30 GHz ($f$ is in GHz), and 200 W/m$^2$ at frequencies from 30 GHz to 300 GHz.

Note that at the transition frequency where the evaluation metric changes from SAR to PD, i.e. 6 GHz for the FCC and 10 GHz for the ICNIRP, the maximum possible radiated power to meet compliance drops about 5.5 dB for the FCC and 6.5 dB for the ICNIRP for a half-wavelength dipole to meet compliance at a separation distance of 2 cm [8]. As a consequence, above 6 GHz for the FCC and 10 GHz for the ICNIRP, the maximum output power is reduced to about 15 dBm and 18 dBm, respectively [8]. Although for IEEE C95.1-2005, this discontinuity is smaller (about 1 dB) at the transition frequency of 3 GHz, due to larger averaging area, it has not yet been adopted by any national regulations. In other words, in order to comply with exposure limits at frequencies above 6 GHz, the maximum radiated power might have to be several dB lower than the power levels used for current mobile technologies. Since the available output power for user devices is critical on the system capacity and coverage, such an inconsistency is undesirable and should be addressed by relevant regulatory authorities to promote the development of future broadband mobile communication networks.

The harmonization of RF exposure limits around the world is highly desired, to provide a consistent protection of all people worldwide, as well as for the wireless industry to serve a worldwide market. Safety determinations for mmWave mobile handsets will raise some novel issues on compliance determinations. First, mmWave handsets will likely to be used close enough to the body, and the resulting fields will be "near-field" rather than "far-field", where reliable PD measurements cannot be obtained. According to the FCC, at frequencies above 6 GHz, reliable PD measurements can normally be made at 5 cm or more from the transmitter [9]. If a device normally operates at a distance closer than 5 cm from persons, PD may be computed using numerical modeling techniques, such as finite-difference time domain (FDTD) or finite element method (FEM) to determine compliance [9]. For example, consider a 60 GHz complementary metal-oxide-semiconductor (CMOS) transceiver for multi-Gb/s wireless communications implemented on a single chip using a 32-element phased-array antenna. It is reasonable to assume that the largest dimension of such an antenna array is $D \approx 10\ mm$ [10][11]. For this example, the far-field distance (Fraunhofer distance $d_{far-field} = 2D^2/\lambda$) is 4 cm. If the RF output power of this transceiver is 100 mW ($P$) and has an antenna gain ($G$) of 10 dB, then for a person located 1 m away from the radiation source ($d$), the peak PD level at the skin surface would be 0.08 W/m$^2$ ($PD_{far-field} = G \cdot P/(4\pi d^2)$). If the distance decreases to 10 cm, which is still in the far-field, the peak radiation level would be 8 W/m$^2$, safely below both the ICNIRP and FCC uncontrolled exposure guidelines of 10 W/m$^2$. If the distance decreases to 5 cm, the peak radiation level would be 32 W/m$^2$, which is above the uncontrolled exposure level of 10 W/m$^2$, but well below both the ICNIRP and FCC occupational/controlled exposure levels of 50 W/m$^2$, and far below the ICNIRP and FCC localized general public/uncontrolled exposure levels of 200 W/m$^2$. For separation distances less than 5 cm, which are normal situations for mobile handsets that are in the pocket or next to the head or hand, numerical modeling rather than direct measurements are needed, thus safety determinations will be complex for antennas of arbitrary geometry and orientation in close vicinity of the highly reflective tissue boundary, and results may vary depending on the methods chosen between different parties conducting compliance evaluations.

MmWave handsets will generally have high gain directional and adaptive antenna arrays [11][12], which causes radiation energy to focus in one or certain directions, leading to increased heating if the main beam points to the human body. Thus, all possible pointing directions of the antenna arrays should be considered to ensure safety, and perhaps a peak value should be used. Moreover, transmission with different amplitude and phase combinations in the adaptive array may result in the creation of constructive/destructive electrical (E) field interference patterns inside the body (although only in the first few millimeters at mmWave frequencies). The power deposition in the body is then roughly proportional to the absolute value squared of the vector addition of the E fields generated by different antenna elements. This capability of E field interactions, particularly with the very small wavelengths involved, means that new quantification methods that account for all possible (and peak) adaptive antenna amplitude and phase configuration should be used [13].

In recent years, the cost of operation of magnetic resonance imaging (MRI) has been decreasing, and MRI-based systems for mapping thermal changes are becoming affordable to wireless manufacturers and regulatory bodies. They provide wideband capabilities, high 3-dimentional resolution, and scan speeds that are unparalleled to the current SAR measurement systems. MRI can accurately measure heating of the skin caused by mmWave radiations. Thus, we propose that temperature-based technology may be a potential method for evaluating safety for future mmWave devices.

II. THE HUMAN BODY EFFECTS ON MMWAVE PROPAGATION

*A. Dielectric Properties of the Skin*

The dielectric properties of the human skin are important for studying mmWave propagation characteristic when radiating sources are in close proximity to the body. Skin consists of two primary layers: an outer epidermis and an underlying dermis, with thicknesses varying in the range of 0.06 to 0.1 mm and 1.2 to 2.8 mm, respectively [13].

The dielectric properties of human skin are obtained from measuring its relative complex permittivity:

$$\varepsilon^* = \varepsilon' - j\varepsilon'' \quad (1)$$

where

$$\varepsilon'' = \frac{\sigma}{2\pi f \varepsilon_0}$$

where $\sigma$ is the conductivity of the material measured in Siemens/meter (S/m), and $\varepsilon_0$ is the permittivity of free space given by $8.85 \times 10^{-12}$ F/m, $f$ is the operating frequency (Hz).

Figs. 1 and 2 show the relative permittivity ($\varepsilon'$) and conductivity ($\varepsilon'' \varepsilon_0 \omega$) of skin versus frequency [3]. The relative permittivity of skin decreases with the increase of frequency, whereas the conductivity of the skin increases with the increase of frequency. The dielectric discrepancies between various studies as seen in Figs. 1 and 2 may be related to the intrinsic differences of measurement methods, and also possibly due to the variations of sample types, such as skin temperature, thickness of different skin layers, etc. It must be noted that many scientific papers make use of the dielectric properties provided by Gabriel et al. at frequencies below 100 GHz, and these data have become widely available through publicly-available online databases [22]. However, these data reflect natural variability in structure and composition of the biological tissues [16]. In order to reasonably predict the effects of the human body on the propagation and absorption of mmWave signals, further dielectric measurements on human skin as well as other body tissues are needed to develop accurate tissue models for mmWave propagation prediction in the presence of humans.

Table I shows the relative complex permittivity ($\varepsilon'$) at 28, 60 and 73 GHz (popular frequencies for mmWave applications [4][10][13][21]) using different skin models.

*B. Reflection and Transmission at the Surface of the Skin*

Since mmWave wavelengths are very short compared with the size of the human body, it is reasonable to model the human skin as a semi-infinite flat surface by considering a mmWave band plane wave illuminating the skin surface. The behavior of an arbitrary wave incident at the skin surface can be studied by considering two distinct cases, parallel polarization (the E-field is parallel to the plane of incidence) and perpendicular polarization (the E-field is perpendicular to the plane of incidence), as shown in Fig. 3. The subscripts *i*, *r*, *t* refer to the incident, reflected and transmitted fields, respectively. The *plane of incident* is defined as the plane containing the incident, reflected, and transmitted rays [23]. The reflection coefficients of parallel and perpendicular polarizations at the boundary of air and skin are given by [23]:

$$R_\parallel = \left| \frac{-\varepsilon^* cos\theta_i + \sqrt{\varepsilon^* - sin^2\theta_i}}{\varepsilon^* cos\theta_i + \sqrt{\varepsilon^* - sin^2\theta_i}} \right| \quad (2)$$

$$R_\perp = \left| \frac{cos\theta_i - \sqrt{\varepsilon^* - sin^2\theta_i}}{cos\theta_i + \sqrt{\varepsilon^* - sin^2\theta_i}} \right| \quad (3)$$

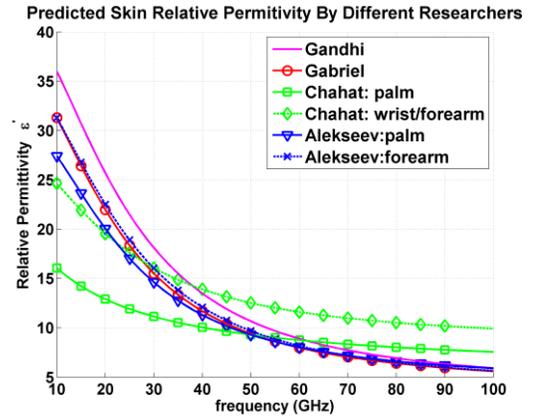

Fig. 1. Predicted skin relative permittivity according to model parameters presented by several researchers from 10 GHz to 100 GHz [3].

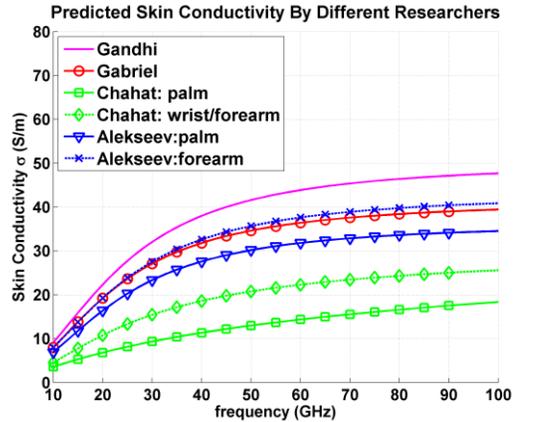

Fig. 2. Predicted skin relative conductivity according to model parameters presented by several researchers from 10 GHz to 100 GHz [3].

TABLE I
RELATIVE COMPLEX PERMITTIVITY AT 28, 60 AND 73 GHZ USING SKIN MODELS DEVELOPED BY DIFFERENT RESEARCHERS

| Skin Models | $f$ (GHz) | | |
|---|---|---|---|
| | 28 | 60 | 73 |
| Gandi [18] | 19.3 - j19.5 | 8.9 - j13.1 | 7.4 - j11.2 |
| Gabriel [15][16][17] | 16.6 - j16.6 | 8.0 - j10.9 | 6.8 - j9.3 |
| Chahat (palm) [20] | 11.4 - j5.7 | 8.7 - j4.3 | 8.2 - j3.9 |
| Chahat (wrist/forearm) [20] | 16.6 - j9.4 | 11.6 - j6.7 | 10.8 - j5.8 |
| Alekseev (palm) [19] | 15.5 - j14.2 | 8.0 - j9.5 | 7.0 - j8.2 |
| Alekseev (forearm) [19] | 17.1 - j16.8 | 8.2 - j11.3 | 6.9 - j9.7 |

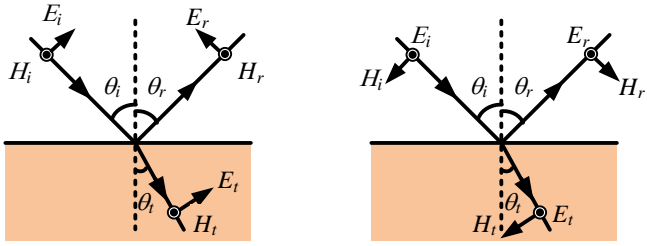

(a) Parallel polarization  (b) Perpendicular polarization

Fig. 3. Parallel and perpendicular polarizations for calculating the reflection coefficients at the air and skin interface.

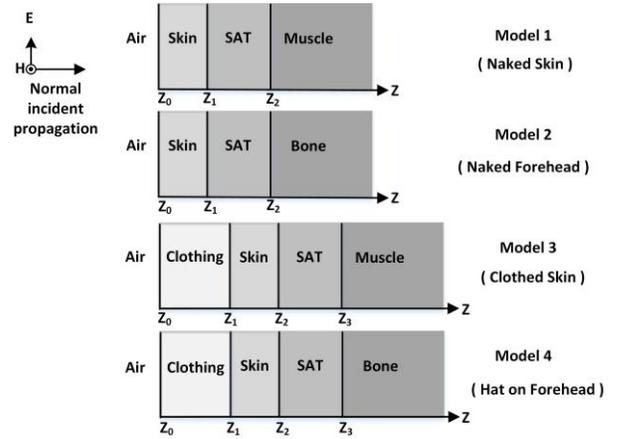

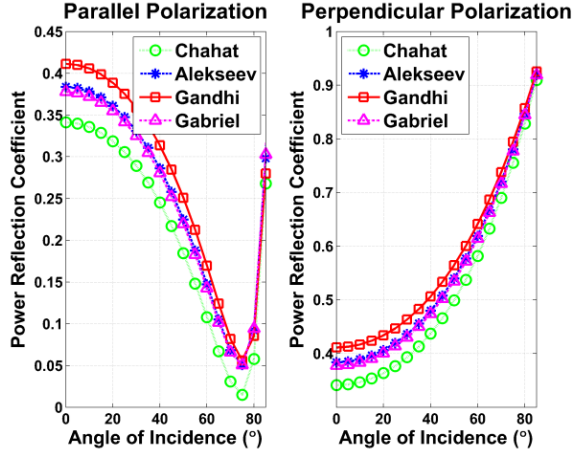

Fig. 4. Power reflection coefficients at the air/skin interface at 60 GHz using different skin model parameters parallel polarization (left) and perpendicular polarization (right).

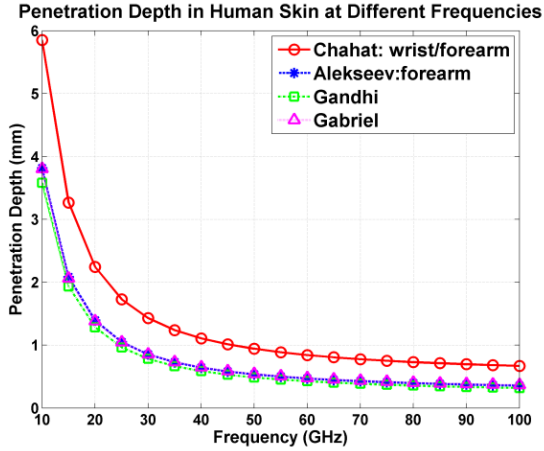

Fig. 5. The penetration depth in the human skin with the increase of exposure frequencies using different skin models [3].

The power reflection coefficient and power transmission coefficient are $R_\parallel^2$ (or $R_\perp^2$) and $1 - R_\parallel^2$ (or $1 - R_\perp^2$), respectively.

Fig. 4 shows the power reflection coefficients at the air and skin interface at 60 GHz for parallel and perpendicular polarized components using various skin model parameters developed by the aforementioned researchers. The results reveal that 34%-42% of the normal incident power is reflected at the skin surface at 60 GHz. The power reflection coefficients vary by 20% when different dielectric model parameters are applied. The Brewster angles where almost all energy is absorbed lie in the range of 65° to 80°.

Fig. 6. Four 1-D human tissue models representing four typical body parts (naked skin, naked forehead, clothed skin, hat on forehead) for the study of heating effects induced by mmWave exposures on the body.

TABLE II
ADOPTED RELATIVE PERMITTIVITY AND CONDUCTIVITY FOR SKIN, SAT, MUSCLE, AND BONE AT 40, 60, 80 AND 100 GHZ

| $f$ GHz | Skin $\varepsilon'$ | $\sigma$ | SAT $\varepsilon'$ | $\sigma$ | Muscle $\varepsilon'$ | $\sigma$ | Bone $\varepsilon'$ | $\sigma$ |
|---|---|---|---|---|---|---|---|---|
| 40 | 11.69 | 31.78 | 5.21 | 6.58 | 18.24 | 43.13 | 4.43 | 6.01 |
| 60 | 7.98 | 36.38 | 4.40 | 8.39 | 12.86 | 52.80 | 3.81 | 7.20 |
| 80 | 6.40 | 38.38 | 3.95 | 9.66 | 10.17 | 58.58 | 3.49 | 8.02 |
| 100 | 5.60 | 39.42 | 3.67 | 10.63 | 8.63 | 62.47 | 3.30 | 8.65 |

TABLE III
ADOPTED MASS DENSITY, THERMAL CONSTANT AND TISSUE THICKNESS FOR SKIN, SAT, MUSCLE, BONE AND BLOOD

| Tissue Properties | Skin | SAT | Muscle | Bone | Blood |
|---|---|---|---|---|---|
| $\rho$ (kg/m$^3$) | 1109 | 911 | 1090 | 1908 | 1050 |
| $c$ (J/kg/°C) | 3391 | 2348 | 3421 | 1313 | 3617 |
| $k$ (W/m/°C) | 0.37 | 0.21 | 0.49 | 0.32 | 0.52 |
| $w$ (mL/kg/min) | 106 | 33 | 37 | 10 | 10000 |
| $Q_m$ (W/m$^3$) [27] | 1620 | 300 | 480 | 0 | 0 |
| Tissue thickness (mm) | 1 | 3 | 31 | 31 | / |

The penetration depth (or skin depth, corresponding to the power density of $1/e^2$ of that transmitted across the surface) of the plane wave in the human body versus frequency using different skin model parameters is shown in Fig. 5. We can see that the penetration depth decreases rapidly with the increase of frequency. Also, more than 90% of the transmitted electromagnetic power is absorbed within the epidermis and dermis layers and little power penetrates further into deeper tissues (although as shown next, the heating of human tissue may extend deeper than the epidermis and dermis layers). Therefore, for the reliable evaluation of mmWave energy distribution in the human body, a single-layer skin model seems to be sufficient.

III. MILLIMETER-WAVE HEATING OF THE SKIN

In this section, the heating effects induced from mmWave exposure are investigated in four one-dimensional (1-D) human tissue models, as shown in Fig. 6, to simulate different body parts. Model 1 represents the tissue layer structure of a naked human body, comprised of skin, subcutaneous adipose tissue (SAT) and muscle. Model 2 illustrates the tissue structure of the naked human forehead. Model 3 simulates the human body covered with clothing and model 4 illustrates the forehead covered with clothing, such as a hat. In order to simplify the problem, we assume a continuous plane wave with radiation frequency $f$ normally incident to the surface of the one-dimensional models of human tissue. The models are infinite on the xy-plane, and semi-infinite along the z-axis.

In each tissue layer, the electric (E) and magnetic (H) fields are:
$$E(z) = E_i^+ e^{-jk_i z} + E_i^- e^{jk_i z}$$
$$H(z) = \frac{E_i^+}{\eta_i} e^{-jk_i z} - \frac{E_i^-}{\eta_i} e^{jk_i z}$$
where $k = \beta - j\alpha = \omega\sqrt{\mu\varepsilon^*}$ and $\eta = \sqrt{\mu/\varepsilon^*}$, where $\omega$ is the angular frequency, $\mu$ is the magnetic permeability and $\varepsilon^*$ is the complex permittivity in the corresponding tissue layer.

The amplitude of the incident wave $E_0^+$ is known for a given radiation PD, while $E_i^- = 0$ in the last layer because the last layer is infinite along the z-axis. The other unknown $E_i^+$ and $E_i^-$ can be found by apply the continuity of both $E$ and $H$ across the interface of different tissue layers.

Most of the theoretical analyses on heat transfer in living tissues are based on the bioheat transfer equation by Pennes [25], which takes into account the effects of blood flow on the temperature distribution in the tissue in terms of volumetrically distributed heat sinks or sources. The one-dimensional version of the bioheat transfer equation is given by [26]:

$$\rho c \frac{\partial T(z)}{\partial t} = k \frac{\partial^2 T(z)}{\partial z^2} - h_b(T(z) - T_{blood}) + Q_m + SAR \cdot \rho \quad (4)$$

where $h_b = \rho_{blood} \cdot w \cdot c_{blood}$ is the heat transfer coefficient, $\rho$ is the mass density in the corresponding tissue layer (kg/m³), $\rho_{blood}$ is the mass density of blood (kg/m³), $c$ is the specific heat capacity in the corresponding tissue layer (J/kg/°C), $c_{blood}$ is the specific heat capacity of blood ((J/kg/°C), k is the thermal conductivity (W/m/°C), w is the perfusion by blood (mL/g/second), T is the tissue temperature (°C), $T_{blood}$ is the blood temperature (°C), $Q_m$ is the heat generated by metabolism (W/m³), and $SAR \cdot \rho$ is the volumetric heat source distributed in the tissue (W/m³) and is given by:

$$SAR \cdot \rho = \frac{\sigma |E(z)|^2}{2\rho} \cdot \rho = \frac{\sigma |E(z)|^2}{2}$$
$$= \frac{\sigma_i}{2}\{[|E_i^+|^2 e^{-2\alpha_i z}] + [|E_i^-|^2 e^{2\alpha_i z}]$$
$$+ [2u_i \cos(2\beta_i z) + 2v_i \sin(2\beta_i z)]\}$$

where $u_i + jv_i = (E_i^+)(E_i^-)^*$.

For the study of steady state temperature elevation, (4) can be further simplified into an ordinary differential equation:
$$0 = k \frac{\partial^2 T}{\partial z^2} - h_b(T - T_{blood}) + Q_m + SAR \cdot \rho$$

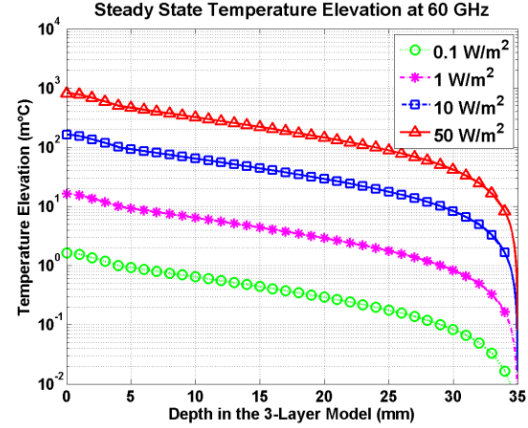

Fig. 7. Steady state temperature elevation at 60 GHz with different incident power densities in naked skin (model 1) [3].

We assume the baseline body temperature before exposure to be $T_s(z)$, the temperature elevation in the human body due to electromagnetic wave exposure can be characterized by $\theta(z) = T(z) - T_s(z)$ and we have:
$$k \frac{\partial^2 \theta}{\partial z^2} - h_b \theta + SAR \cdot \rho = 0$$

The above ordinary differential equation can be solved analytically [26]:
$$\theta(z) = \varphi(z) + \zeta(z) + \xi(z) + \psi(z)$$
where $\varphi(z)$ is the general solution of the corresponding homogeneous equation, $\zeta(z), \xi(z)$ and $\psi(z)$ are the three particular solutions of the corresponding nonhomogeneous equation and they are given by:

$$\varphi(z) = C_A e^{-\sqrt{\frac{h_b}{k}}z} + C_B e^{\sqrt{\frac{h_b}{k}}z}$$
$$\zeta(z) = -\frac{\sigma}{2(4\alpha^2 k - h_b)} |E^+|^2 e^{-2\alpha z}$$
$$\xi(z) = -\frac{\sigma}{2(4\alpha^2 k - h_b)} |E^-|^2 e^{2\alpha z}$$
$$\psi(z) = \frac{\sigma}{2(4\beta^2 k + h_b)} [u \cos 2\beta z + v \sin 2\beta z]$$

$C_A$ and $C_B$ in each tissue layer can be solved by forcing boundary conditions[26] shown below:
a. At the external skin surface:
$$k_1 \frac{\partial T(z)}{\partial z}\Big|_{z=Z_0} = h(T(z_0) - T_{air}) \quad (5)$$
where $h$ is the heat transfer coefficient and is 7 W/m²/°C from the outer skin surface to air and 0 from the outer skin surface to clothing. Note that for models 3 and 4, $Z_0$ should be replaced with $Z_1$.
b. At the other interfaces, continuity of both temperature and heat flux should be satisfied:
$$T(Z_i^-) = T(Z_i^+), \quad k_{i-1}\frac{\partial T(Z_i^-)}{\partial z} = k_i \frac{\partial T(Z_i^+)}{\partial z} \quad (6)$$
c. Finally, the steady state temperature elevation at 35 mm inside the tissue is enforced to be 0 °C. In other words, the steady state temperature at places deeper than 35 mm inside the tissue is equal to the blood temperature.

The tissue properties listed in Table II and Table III have been chosen according to the database developed by Hasgall *et al* [22]. The thickness of the clothing is 1 mm (if not specified) with a relative complex permittivity of 1.6 + j0.06 which is estimated from the complex permittivity of denim measured at 40 GHz [24]. $T_{blood}$ is 37 °C and $T_{air}$ is 23 °C in the simulation.

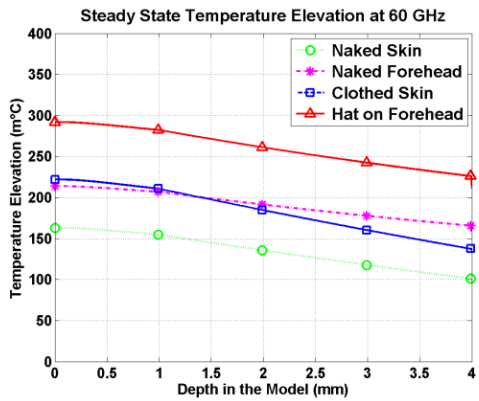

Fig. 8. Steady state temperature elevation due to 10 W/m² at 60 GHz in the four models shown in Fig. 5 from the skin surface to 4 mm in the tissue.

Fig. 7 shows the steady state temperature elevation at 60 GHz with incident power densities of 0.1 W/m² (PD limits for China, Russia, Switzerland, and Italy [3]), 1 W/m², 10 W/m² (FCC and ICNIRP PD restrictions for the general public) and 50 W/m² (FCC/ICNIRP PD restrictions for the occupational group) in naked skin. It can be seen that the steady state temperature elevation is proportional to the incident power densities. When the incident power density is 50 W/m², the temperature elevation at the skin surface is about 0.8 °C, which is below the temperature threshold of 1 °C according to IEEE standards on mmWave radiation guidelines [2][5].

Fig. 8 shows the steady state temperature elevation due to 10 W/m² at 60 GHz in the four models. Naked skin (model 1) produces the least heat since the heat generated in the skin can be dissipated into the air and taken away by the blood flow in the muscle. Thus, the steady state temperature elevation in naked skin is the lowest (only 0.16 °C). While hat on forehead (model 4) generates the most heat since the skin is covered with clothing and the bone lacks blood flow to take away the heat generated, and not allowing thermal conduction into the air or even within the bone. Thus, the steady state temperature elevation at the skin surface of forehead with hat is the highest (0.3 °C). The steady state temperature elevation in naked forehead (model 2) is low in the skin surface but high in the underlying tissues (SAT and bone) compared with clothed skin (model 3). The low temperature elevation in the skin surface of naked forehead comes from the low heat source distribution (SAR· $\rho$ distributions) in the skin as well as the thermal conduction into the air, while the high steady state temperature elevation in the underlying tissues comes from the poor heat conduction capability of bone.

Fig. 9 shows the effects of clothing thickness on the power transmission coefficients at the air/clothing interface and clothing/skin interface. Both power transmission coefficients

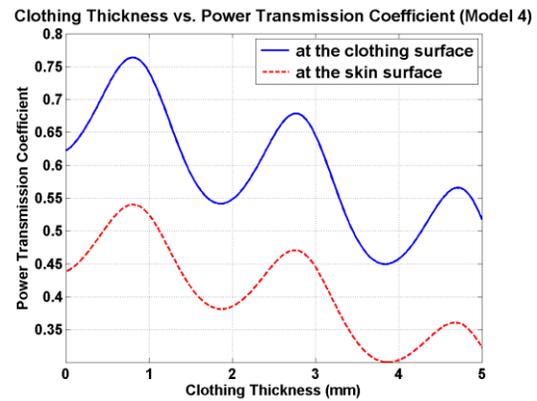

Fig. 9. The dependence of clothing thickness upon the power transmission coefficient at 60 GHz with an incident power density of 10 W/m² for hat on forehead (model 4).

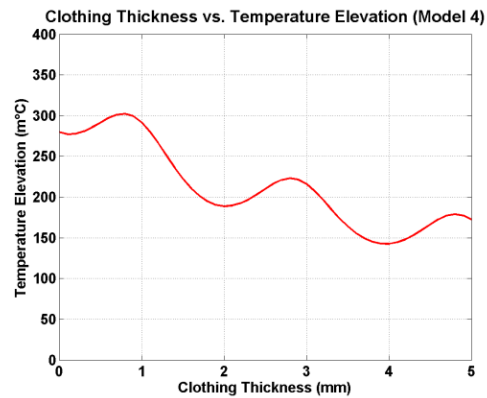

Fig. 10. The dependence of clothing thickness upon the steady state temperature elevation at the skin surface at 60 GHz with an incident power density of 10 W/m² for hat on forehead (model 4).

are calculated with respect to the incident power at the clothing surface using the following equations:

$$\frac{P_{a\_c}}{P_i} = 1 - |R_0|^2 \quad (7)$$

$$\frac{P_{c\_s}}{P_i} = (1 - |R_0|^2)(1 - |R_1|^2)e^{-2\alpha_1 d_c} \quad (8)$$

where $P_{a\_c}$ and $P_{c\_s}$ are the transmitted power at the air/clothing interface and clothing/skin interface, $R_0$ and $R_1$ are the reflection coefficients at the air/clothing interface and clothing/skin interface, $\alpha_1$ is the attenuation constant of clothing and $d_c$ is the thickness of clothing. At 60 GHz, the wavelength in the clothing is about 3.95 mm ($\varepsilon^*$=1.6 + j0.06). The local peak power transmissions happen every half wavelength and the overall power transmission decreases due to the attenuation of the clothing. When the clothing thickness is less than 1 mm, the clothing may act like an impedance transformer resulting in the enhancement of the power transmitted into the skin [18]. Fig. 10 shows the corresponding temperature elevation due to the increase of clothing thickness. Local peak temperature elevations can be observed every half wavelength.

From Figs. 8 to 10, we can see that the steady state temperature elevations at different body locations may vary even when the intensities of electromagnetic wave radiations are the same. This is obvious since PD does not consider the reflection or transmission of mmWave energy across boundaries. Hence, PD is not likely to be as useful as SAR for assessing safety, especially in the near-field. We propose that temperature-based technique using MRI may be considered an acceptable dosimetric quantity for demonstrating safety [3].

IV. CONCLUSION

In this paper, global regulations for mmWave exposure were presented, and an example of power levels and current regulations for a 60 GHz device was provided. The importance of a sound dielectric database was shown by comparing the predicted power reflection and transmission coefficient in the skin using different skin dielectric models. At 60 GHz, the power reflection coefficient may vary between 34% and 42% at the air/skin interface for the normal incidence due to variations of dielectric parameters. The analyses of penetration depth show that more than 90% of the transmitted power is absorbed in the epidermis and dermis layer, suggesting that a single-layer skin model is sufficient for a reliable electromagnetic evaluation in the human body.

However, for thermal modeling, a multi-layer skin model is preferred since the heat at the surface must be conducted through skin and underlying tissues (e.g., SAT and muscle). We used four one-dimensional models of the human tissue to illustrate the effects of thermal heating and electromagnetic penetration into skin. The dependence of clothing thickness upon the power transmission coefficient and steady state temperature elevation was studied. We have suggested the use of temperature elevation in the human head or body as a valid compliance evaluation method for mmWave exposure, since temperature changes in the human body have a more straightforward relationship with safety than power density. Measurements or simulations of temperature increase are currently acceptable for showing compliance to limits on exposure to radio frequency energy in MRI [28].